\documentclass[twocolumn,showpacs,preprintnumbers]{revtex4}
\usepackage{amsfonts}
\usepackage{amssymb}
\usepackage{amsmath}
\usepackage{graphicx}
\usepackage{dcolumn}
\usepackage{bm}

\begin{document}

\title{Solution of the two-mode quantum Rabi model using extended squeezed
states}
\author{Liwei Duan$^{1}$, Shu He$^{1}$, D. Braak$^{2,3,\dagger}$, and
Qing-Hu Chen$^{1,4,*}$}

\address{
$^{1}$ Department of Physics, Zhejiang University, Hangzhou 310027,
 China \\
$^{2}$  EP VI and Center for Electronic Correlations and Magnetism, University of Augsburg, 86135 Augsburg, Germany\\
$^{3}$ Center for Correlated Matter, Zhejiang University, Hangzhou 310058,
China \\
$^{4}$  Collaborative Innovation Center of Advanced Microstructures,
Nanjing 210093, China
 }\date{\today }

\begin{abstract}
The two-mode quantum Rabi model with bilinear coupling is studied using
extended squeezed states. We derive $G$-functions for each Bargmann index $q$%
. They share a common structure with the $G$-function of the one-photon
and two-photon quantum Rabi models.
The regular spectrum is given by zeros of the $G$-function while the
conditions for the presence of doubly degenerate (exceptional) eigenvalues
are obtained in closed form through the lifting property. The simple
singularity structure of the $G$-function allows to draw conclusions about
the distribution of eigenvalues along the real axis and to understand the
spectral collapse phenomenon when the coupling reaches a critical value.
\end{abstract}

\maketitle

\section{Introduction}

The Quantum Rabi model (QRM) describes a two-level system (qubit) coupled to
a cavity electromagnetic mode (an oscillator)~\cite{Rabi,Jaynes-Cummings}, a
minimalist paradigm of matter-light interactions with applications in
numerous fields ranging from quantum optics to quantum information science
and condensed matter physics. Its Hamiltonian reads,
\begin{equation}
H_R=\Delta\sigma _z + \omega a^{\dagger }a +g(a^{\dagger }+a)\sigma _x.
\label{rabimodel}
\end{equation}
The Pauli matrices $\sigma_{x,z}$ describe the two-level system and $a$,($%
a^\dagger$) denote the annihilation (creation) operators of the bosonic
mode. Simple as it appears, the solution of the QRM is nontrivial and it was
not clear whether it can be obtained analytically. Recently it was shown
that the QRM is not only exactly solvable but integrable~\cite{Braak}, using
Bargmann-space methods~\cite{Bargmann}. A function $G_R(E)$ was derived,
whose zeros yield the so-called regular spectrum, i.e. $G_R(E_n)=0$ entails $%
E_n\in \text{spec}(H_R)$.
This $G$-function can be written explicitly in terms of confluent Heun
functions~\cite{Slavyanov}. $G_R(E)$ was then recovered with the simpler
formalism of extended coherent states~\cite{Chen2012}. These results have
stimulated extensive research in the QRM and related models~\cite%
{twophoton,Moroz,Gardas,Braak2013,Maciejewski,Zhong,wang2014,Fanheng,Tomka,Peng,Chilingaryan,Chennext,Batchelor}%
.

It is well known in quantum optics~\cite{book97} that two-mode squeezed
states are important since several devices produce correlated light at two
frequencies. Recently, a quantum memory for light, a key element for the
realization of future quantum information networks, was constructed
employing a set of displaced two-mode squeezed states with an
unconditionally high fidelity that exceeds the classical benchmark~\cite%
{memory}. Two-mode squeezed states have been prepared also in cavity
optomechanics via reservoir engineering~\cite{exptwomode}. There are many
hints to potential applications of a single qubit coupled simultaneously to
two different light modes in quantum information technology, e.g. for the
implementation of fast beam splitters~\cite{Rodriguez}. Such a system may
also be realized in various solid state devices~\cite%
{Sillanpaa,2qubits,Dicalo}, and especially in circuit QED~\cite{Niemczyk}.

One analytically tractable form of the two-mode generalization of the QRM
couples the qubit bilinearly to the two cavity modes, thereby creating
squeezed states. This model will be studied in the next section. We derive a
$G$-function in analogy to the single-mode case and obtain regular and
exceptional spectra analytically. Most important, we can derive the overall
features of the spectrum from the pole structure of the $G$-function alone,
without the need for numerical evaluation, and explain in this way the
collapse of the discrete spectrum to a continuum which happens in this model
as in the QRM with quadratic coupling to a single mode~\cite%
{twophoton,collapse}.

\section{Two-mode Rabi model}

The Hamiltonian for the two-mode QRM~\cite{Zhangyz} reads,
\begin{equation}
H_{tm}=\Delta \sigma _z+\omega \left( a_1^{\dagger
}a_1+a_2^{\dagger}a_2\right) +g\left( a_1^{\dagger }a_2^{\dagger
}+a_1a_2\right) \sigma _x,  \label{model}
\end{equation}
where $g$ is coupling strength and $a_i^{\dagger }$ and $a_i$ are the
creation and annihilation operators for the quantized fields in two cavities
$i=1,2$.
The energy scale is defined by setting $\omega =1$.

A rotation around the $y$-axis through the angle $\frac \pi 2$ yields the
Hamiltonian in spin-boson form,
\begin{widetext}
\begin{equation}
H_{sb}=\left(
\begin{array}{ll}
\left( a_1^{\dagger }a_1+a_2^{\dagger }a_2\right) +g\left( a_1^{\dagger
}a_2^{\dagger }+a_1a_2\right) & \;\;\;\;\;\;\;\;\;\;\;\;\;\;\;\;- \Delta \\
\;\;\;\;\;\;\;\;\;\;\;\;\;\;\;\;-\Delta & \left( a_1^{\dagger
}a_1+a_2^{\dagger }a_2\right) -g\left( a_1^{\dagger }a_2^{\dagger
}+a_1a_2\right)%
\end{array}
\right),  \label{Hamiltonian}
\end{equation}
\end{widetext}
which will be convenient in the following.

\textsl{Symmetries:} The Hamiltonian \eqref{model} exhibits a continuous $%
U(1)$-symmetry similar to the Jaynes-Cummings model~\cite{Jaynes-Cummings};
it is invariant under the transformation
\begin{equation}
\begin{array}{cccccc}
a_{1} & \rightarrow  & e^{i\phi }a_{1}, & \quad a_{1}^{\dagger } &
\rightarrow  & e^{-i\phi }a_{1}^{\dagger } \\
a_{2} & \rightarrow  & e^{-i\phi }a_{2}, & \quad a_{2}^{\dagger } &
\rightarrow  & e^{i\phi }a_{2}^{\dagger }%
\end{array}%
\end{equation}%
for arbitrary $0\leq \phi <2\pi $. This symmetry is generated by the
operator $\hat{C}=a_{1}^{\dagger }a_{1}-a_{2}^{\dagger }a_{2}$, which
commutes with $H_{tm}$. Accordingly, there exist infinitely many invariant
subspaces $\mathcal{H}_{m}$ labeled by the eigenvalue $m\in
%TCIMACRO{\U{2124} }%
%BeginExpansion
\mathbb{Z}
%EndExpansion
$ of $\hat{C}$. Confinement to a single $\mathcal{H}_{m}$ effectively
eliminates one of the bosonic modes. Conforming to the usual terminology~%
\cite{Zhangyz}, we set $m=2q-1$ and call $q$ the Bargmann index of $\mathcal{%
H}_{m}$. In the following we consider $m\geq 0$ ($m<0$ exchanges modes 1 and
2), i.e. $q$ runs over positive multiples of 1/2. Besides the continuous
symmetry acting only in the bosonic subspace, $H_{tm}$ possesses also a
discrete symmetry $\hat{P}$ which takes the form $\hat{P}=\sigma _{z}\exp
(i\pi a_{2}^{\dagger }a_{2})$. We have $\hat{P}H_{tm}%
\hat{P}^{\dagger} =H_{tm}$ and $\hat{P}^{2}={1\!\!1}$. $\hat{P}$ generates therefore a $%
%TCIMACRO{\U{2124} }%
%BeginExpansion
\mathbb{Z}
%EndExpansion
_{2}$-symmetry (parity). It has eigenvalues $\pm 1$, entailing a further
decay of $\mathcal{H}_{2q-1}$ into two invariant subspaces $\mathcal{H}%
_{2q-1}^{\pm }$, labeled by the parity quantum number. These two symmetries
together render the two-mode QRM \eqref{model} integrable according to the
level-labeling criterion~\cite{Braak, Batchelor}.

\textsl{Bogoliubov transformations:} In analogy to the two-photon QRM~\cite%
{Chen2012}, we may perform a bosonic Bogoliubov transformation to new
operators $b_j,b_j^\dagger$, $j=1,2$, which in the present case mixes the
two modes,
\begin{eqnarray*}
b_1 &=&ua_1+va_2^{\dagger },\quad b_1^{\dagger }=ua_1^{\dagger }+va_2, \\
b_2 &=&ua_2+va_1^{\dagger },\quad b_2^{\dagger }=ua_2^{\dagger }+va_1,
\end{eqnarray*}
%to generate two new bosonic operators if $u^2-v^2=1$.
The $b_j$ satisfy bosonic commutation relations if $u^2-v^2=1$. Because
\begin{equation*}
\hat{C}\equiv a_1^\dagger a_1 -a_2^\dagger a_2=\hat{C}_b\equiv b_1^\dagger
b_1-b_2^\dagger b_2,
\end{equation*}
this transformation leaves each subspace $\mathcal{H}_{2q-1}$ invariant.
%of the boson operators $\left[ b,b^{\dagger
%}\right] =1$.
With the choice
\begin{equation}
u=\sqrt{\frac{1+\beta }{2\beta }},\quad v=\sqrt{\frac{1-\beta }{2\beta }},
\label{uv}
\end{equation}
where $\beta =\sqrt{1-g^2}$, the upper left block of the Hamiltonian matrix %
\eqref{Hamiltonian} becomes
\begin{equation*}
H^{11}_b=\beta \left( b_1^{\dagger }b_1+b_2^{\dagger }b_2+1\right) -1,
\end{equation*}
which no longer contains the squeezing terms.

Alternatively, we could transform to operators $c_j,c_j^\dagger$ via
\begin{eqnarray*}
c_1 &=&ua_1-va_2^{\dagger },\quad c_1^{\dagger }=ua_1^{\dagger }-va_2, \\
c_2 &=&ua_2-va_1^{\dagger },\quad c_2^{\dagger }=ua_2^{\dagger }-va_1,
\end{eqnarray*}
and $u,v$ from Eq.~\eqref{uv}. The lower right block of \eqref{Hamiltonian}
becomes now

\begin{equation*}
H^{22}_c=\beta \left( c_1^{\dagger }c_1+c_2^{\dagger }c_2+1\right) -1,
\end{equation*}
and is free from squeezing terms.

In terms of $b$-operators, the Hamiltonian \eqref{Hamiltonian} reads,
\begin{equation}
H_{b}=\left(
\begin{array}{ll}
H_{b}^{11} & -\Delta \\
-\Delta & H_{b}^{22}%
\end{array}%
\right) .  \label{H_b}
\end{equation}%
with
\begin{eqnarray*}
H_{b}^{22} &=&\frac{1+g^{2}}{\beta }\left( b_{1}^{\dagger
}b_{1}+b_{2}^{\dagger }b_{2}\right) \\
&&-\frac{2g}{\beta }\left( b_{1}^{\dagger }b_{2}^{\dagger
}+b_{1}b_{2}\right) -\frac{1}{\beta }\left( \beta +\beta ^{2}-2\right) .
\end{eqnarray*}%
Only the lower diagonal block contains terms of the form $b_{1}^{\dagger
}b_{2}^{\dagger }$, $b_{1}b_{2}$.

We confine the analysis now to the space $\mathcal{H}_{2q-1}$ and define
three sets of mutually orthogonal vectors $\{|n\rangle _{s}^{q}\}$, each
forming an ONB for $\mathcal{H}_{2q-1}$,
\begin{equation}
|n\rangle _{s}^{q}=\frac{\left( s_{1}^{\dagger }\right) ^{n+2q-1}\left(
s_{2}^{\dagger }\right) ^{n}}{\sqrt{\left( n+2q-1\right) !n!}}|0\rangle
_{s},\quad n=0,1,2,\ldots
\end{equation}%
for $s=a,b,c$. $|0\rangle _{s}$ denotes the normalized vacuum state~\cite%
{squeezed} belonging to the operators of type $s$, satisfying $%
s_{1}|0\rangle _{s}=s_{2}|0\rangle _{s}=0$ in $\mathcal{H}$. The
corresponding vacuum state in $\mathcal{H}_{2q-1}$ is $|0\rangle
_{s}^{q}=((2q-1)!)^{-1/2}{(s_{1}^{\dagger })}^{2q-1}|0\rangle _{s}.$ %
We call $|n\rangle _{s}^{q}$ the two-mode extended squeezed states (ESS) and
the set $\{|n\rangle _{s}^{q}\}$ the $s$-basis for $\mathcal{H}_{2q-1}$.
Later we shall use the overlaps %

\begin{eqnarray}
\phantom{0}_{a}^{q}\langle 0|n\rangle _{b}^{q} &=&(-1)^{n}\phantom{!}%
_{a}^{q}\langle 0|n\rangle _{c}^{q}  \notag \\
&=&\left( 1-\frac{v^{2}}{u^{2}}\right) ^{q}\left( \frac{v}{u}\right) ^{n}%
\sqrt{\frac{\left( n+2q-1\right) !}{n!\left( 2q-1\right) !}},
\label{bc_coef}
\end{eqnarray}%
between the vacuum of the $a$-basis and arbitrary states in the $b$- resp. $%
c $-basis.

\vspace{5mm}
$G$\textsl{-functions}:
We drop the index $q$ from now on and consider the Schr\"{o}dinger equation $%
H_{sb}|\psi \rangle =E|\psi \rangle $. We make an ansatz for $|\psi \rangle $
using the $b$-basis
\begin{equation}
\left\vert \psi \right\rangle _{b}=\left( \
\begin{array}{l}
\sum_{n=0}e_{n}\sqrt{\left( n+2q-1\right) !n!}\left\vert n\right\rangle _{b}
\\
\sum_{n=0}f_{n}\sqrt{\left( n+2q-1\right) !n!}\left\vert n\right\rangle _{b}%
\end{array}%
\right) .  \label{wave_b}
\end{equation}%
With Eq.~\eqref{H_b} we obtain the following recurrence relation for the
coefficients $e_{n},f_{n}$,
\begin{equation}
e_{n}=\frac{\Delta }{2\beta \left( n+q\right) -1-E}f_{n},  \label{e-f}
\end{equation}%
\begin{eqnarray}
\Delta e_{n} &=&\left( \frac{1+g^{2}}{\beta }2\left( n+q\right) -1-E\right)
f_{n}  \notag \\
&&-\frac{2g}{\beta }\left[ f_{n-1}+\left( n+1\right) \left( n+2q\right)
f_{n+1}\right] .  \label{recur-1}
\end{eqnarray}%
Note that Eq.~(\ref{e-f}) expresses $e_{n}$ in terms of $f_{n}$, so that a
linear three-term recurrence relation is obtained for the $f_{n}$,
\begin{eqnarray}
f_{n+1} &=&\frac{\left( 1+g^{2}\right) \left( n+q\right) -\beta ^{2}\left(
q+x\right) +\frac{\Delta ^{2}}{4\left( x-n\right) }}{g\left( n+1\right)
\left( n+2q\right) }f_{n}  \notag \\
&&-\frac{f_{n-1}}{\left( n+1\right) \left( n+2q\right) }  \label{recurrence}
\end{eqnarray}%
where we have defined the spectral parameter $x=(E+1)/(2\beta )-q$.

The initial conditions for the recurrence \eqref{recurrence} are $f_{-1}=0$
and $f_{0}=r_{b}$, where $r_{b}$ may be used to normalize $|\psi \rangle
_{b} $ if this is possible. Of course, $|\psi \rangle _{b}$ will be only
normalizable if $E$ coincides with an eigenvalue of $H_{sb}$.

The formal solution $|\psi \rangle $ of $H_{sb}|\psi \rangle =E|\psi \rangle
$ can also be expanded in the $c$-basis as
\begin{equation}
\left\vert \psi \right\rangle _{c}=\left( \
\begin{array}{l}
\sum_{n=0}\left( -1\right) ^{n}w_{n}\sqrt{\left( n+2q-1\right) !n!}%
\left\vert n\right\rangle _{c} \\
\sum_{n=0}\left( -1\right) ^{n}z_{n}\sqrt{\left( n+2q-1\right) !n!}%
\left\vert n\right\rangle _{c}%
\end{array}%
\right) .  \label{wave_c}
\end{equation}%
It turns out that the $w_{n}$ and $z_{n}$ satisfy the same relations %
\eqref{e-f} and \eqref{recur-1} as $f_{n}$ and $e_{n}$.
%we can easily get the exactly same relations for $z_n^{(q)}$ and $w_n^{(q)}$
%as Eq. (\ref{recurrence}). So we have
Thus,
\begin{equation*}
w_{n}=rf_{n},\quad z_{n}=re_{n},
\end{equation*}%
with some constant $r$. Now we assume the state $|\psi \rangle $ to be the
unique eigenvector of $H_{sb}$ belonging to energy $E$, i.e. $|\psi \rangle
_{b}=|\psi \rangle _{c}$.
It follows,
\begin{eqnarray}
&&\sum_{n=0}e_{n}\sqrt{\left( n+2q-1\right) !n!}\left\vert n\right\rangle
_{b}  \notag \\
&=&r\sum_{n=0}\left( -1\right) ^{n}f_{n}\sqrt{\left( n+2q-1\right) !n!}%
\left\vert n\right\rangle _{c},  \label{Eq_ef1} \\
&&\sum_{n=0}f_{n}\sqrt{\left( n+2q-1\right) !n!}\left\vert n\right\rangle
_{b}  \notag \\
&=&r\sum_{n=0}\left( -1\right) ^{n}e_{n}\sqrt{\left( n+2q-1\right) !n!}%
\left\vert n\right\rangle _{c}.  \label{Eq_ef2}
\end{eqnarray}%
Projecting both sides of Eqs.~\eqref{Eq_ef1}, \eqref{Eq_ef2} onto the vacuum
of the $a$-basis $|0\rangle _{a}^{q}$ yields %\begin{equation}
\begin{equation}
\begin{array}{ccc}
\Gamma _{e}(x) & = & r\Gamma _{f}(x), \\
\Gamma _{f}(x) & = & r\Gamma _{e}(x),%
\end{array}
\label{G-cond}
\end{equation}%
with
\begin{equation}
\begin{array}{ccc}
\Gamma _{f}(x) & = & \sum_{n=0}f_{n}(x)(n+2q-1)!\left( \frac{v}{u}\right)
^{n}, \\
\Gamma _{e}(x) & = & \sum_{n=0}e_{n}(x)(n+2q-1)!\left( \frac{v}{u}\right)
^{n},%
\end{array}%
\end{equation}%
where Eq. (\ref{bc_coef}) has been used. Eliminating $r$ from Eq.~%
\eqref{G-cond} leads to $\Gamma _{f}(x)\pm \Gamma _{e}(x)=0$. With Eq.~%
\eqref{e-f}, we can define two $G$-functions,
\begin{widetext}
\begin{equation}
G_{\pm }^{q}\left( x\right) \equiv \Gamma _{f}(x)\pm \Gamma
_{e}(x)=\sum_{n=0}^{\infty }f_{n}^{q}(x)
\left[ 1\pm \frac{\Delta }{2\beta \left( n-x\right) }\right] \left(
n+2q-1\right) !\left( \frac{v}{u}\right) ^{n},  \label{G-function}
\end{equation}
\end{widetext}
each corresponding to a subspace of $\mathcal{H}_{2q-1}$ with fixed parity.
If $x_{n}^{\pm }$ satisfies $G_{\pm }^{q}(x_{n}^{\pm })=0$, we conclude that
$E_{n}^{\pm }=2\beta (x_{n}^{\pm }+q)-1$ is a non-degenerate eigenvalue of $%
H_{tm}$.

The reason for the equivalence of the recurrence relations for $%
\{e_{n},f_{n}\}$ and $\{z_{n},w_{n}\}$, which made the foregoing arguments
possible, is the discrete $%
%TCIMACRO{\U{2124} }%
%BeginExpansion
\mathbb{Z}
%EndExpansion
_{2}$-symmetry of the two-mode QRM, just as for the single-mode, linearly
coupled model~\cite{Braak,Chen2012}. However, while the representation in
the Bargmann space provides a mathematically rigorous justification of the
method in the latter case~\cite{Braak-ann}, the present analysis using ESS
needs to be rederived by imposing the proper normalization conditions on the
eigenstates of $H_{sb}$ in the space $\mathcal{H}_{2q-1}$, which will be the
subject of future work.

\begin{figure}[tbp]
\includegraphics[width=8cm]{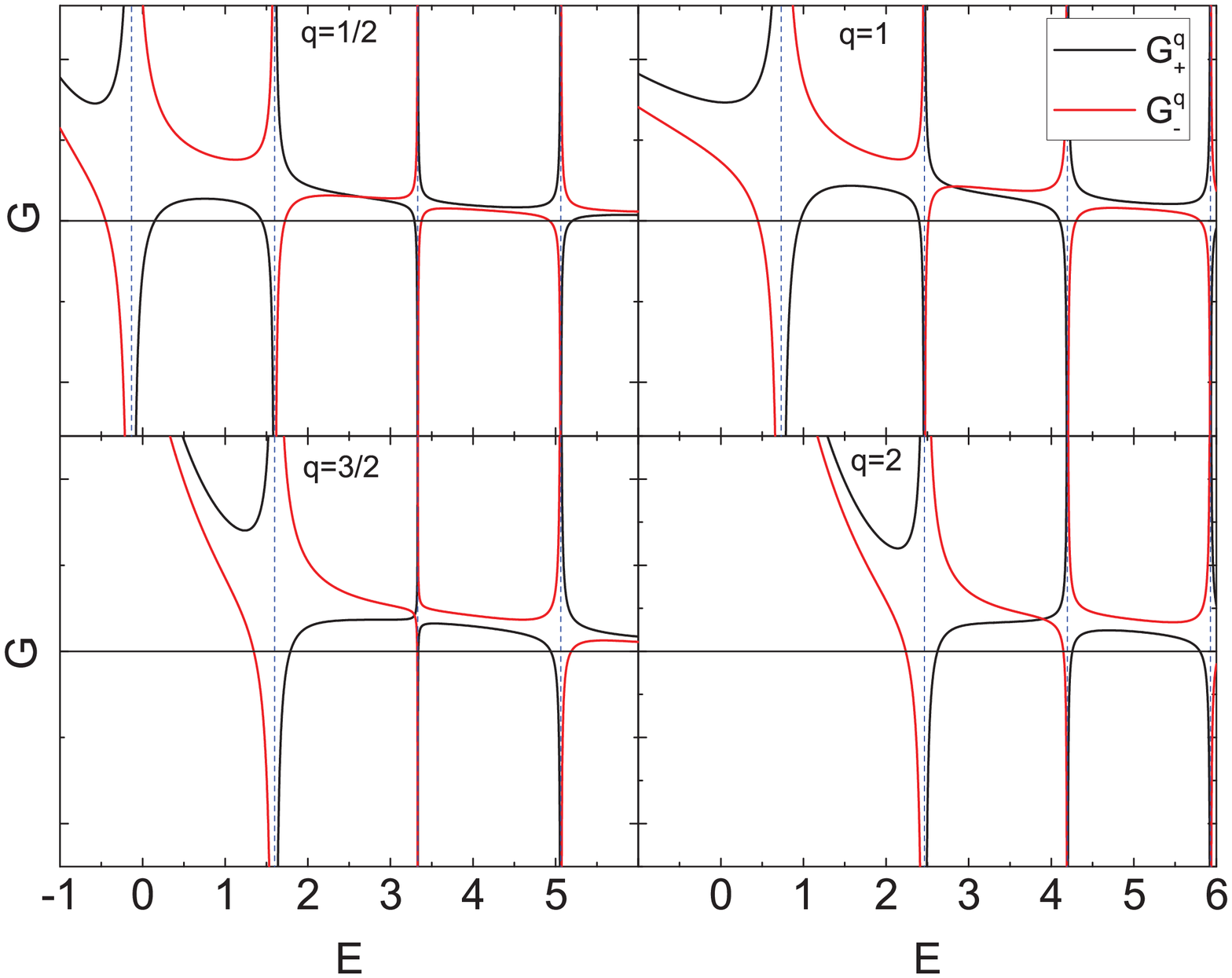} %
\includegraphics[width=8cm]{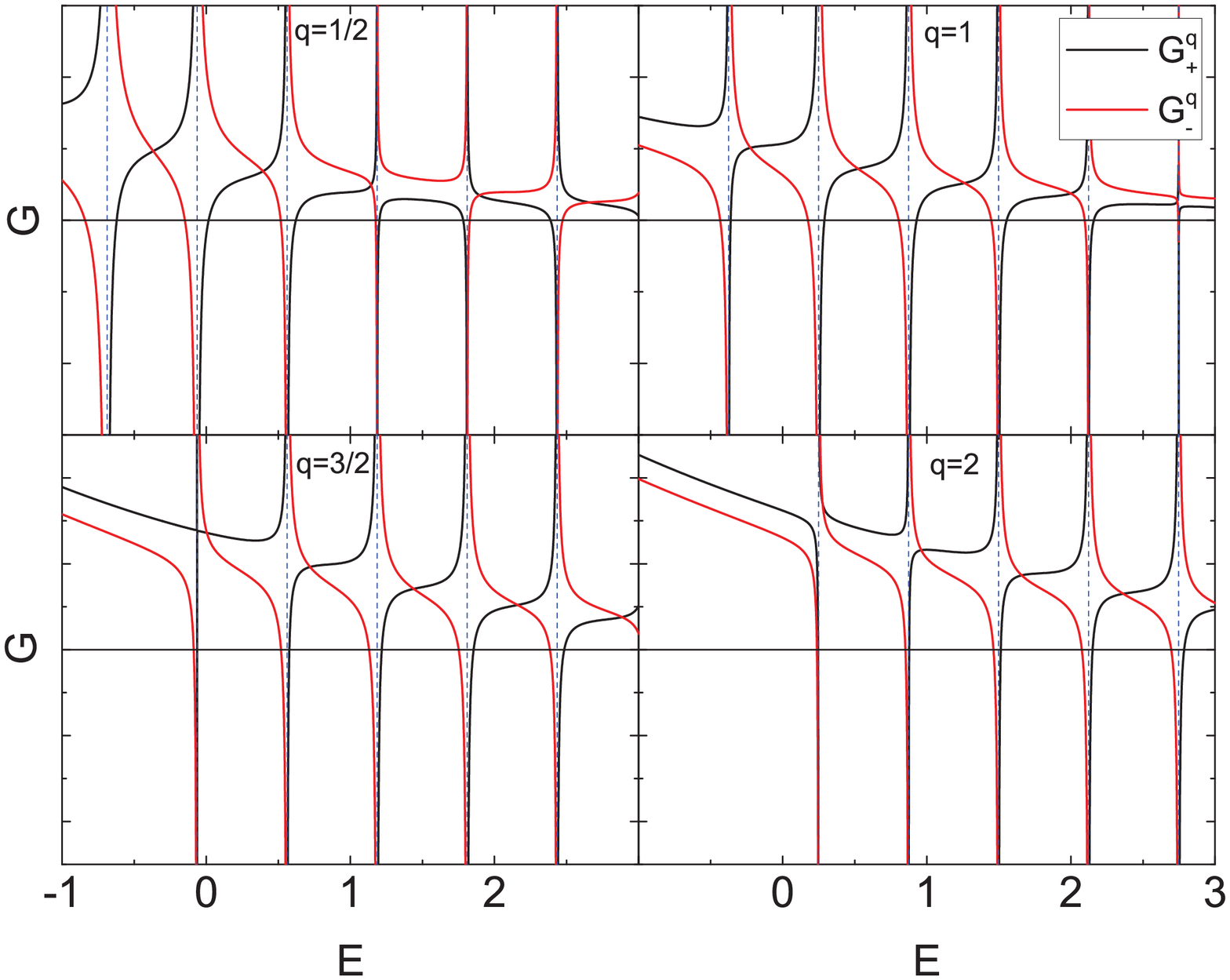}
\caption{(Color online) $G(x)$ for the two-mode QRM at $\Delta =0.35$, $g=0.5
$ (upper panel) and $g=0.95$ (lower panel) for $q=1/2,1,3/2$ and $2$. Arbitrary scale
is employed in the $G$-axis for the visibility of the curves.  The
black (red) curves denote positive (negative) parity.}
\label{gfunction}
\end{figure}

\begin{figure}[tbp]
\includegraphics[width=8cm]{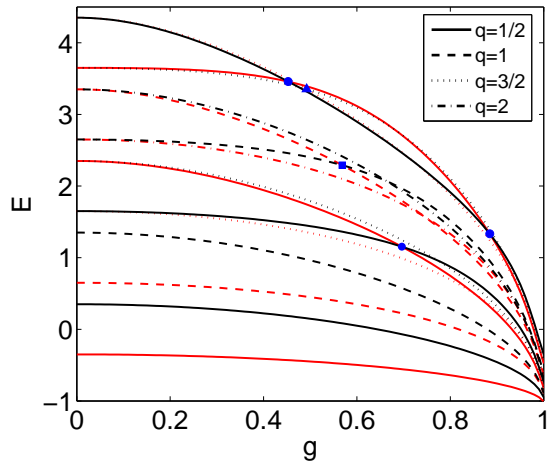}
\caption{(Color online) The spectrum and the isolated exceptional solutions
for the two-mode QRM at $\Delta =0.35$ for $q=1/2,1,3/2$ and $2$. The black
(red) one denotes positive (negative) parity.}
\label{spectrum05}
\end{figure}

\textsl{Regular spectrum}: We plot the $G$-functions in Fig.~\ref{gfunction}
for the case of $\Delta =0.35$ and $g=0.5$ and $0.95$ for some values of $q$%
. The zeros $x^\pm_n$ of the functions $G_\pm^q(x)$ yield the eigenvalues $%
E_n^\pm$ of the regular spectrum, depending on the parameters $q,\Delta,g$.
Fig.~\ref{spectrum05} displays the spectrum of the two-mode QRM as a function
of $g$ for qubit splitting $\Delta =0.35$. We find complete agreement with
numerical diagonalization in truncated Hilbert spaces of sufficiently high
dimension.

This is to be expected, because the three-term recurrence relation Eq.~%
\eqref{recurrence} allows for minimal and dominant solutions like the QRM~%
\cite{Braak}. One can show~\cite{Zhangyz} that only the unique minimal
solution is normalizable and corresponds therefore to a point of the
discrete spectrum if the energy is chosen such that it matches the initial
conditions of \eqref{recurrence}. It is therefore possible to derive a $G$%
-function for the two-mode QRM based on continued fractions, which is
equivalent to truncate the model on a finite-dimensional Hilbert space~\cite%
{Braak-contfrac}. However, all qualitative information about the features of
the spectrum is lost in this approach as the singularity structure of the
continued fraction is not known. To the contrary, the $G$-functions %
\eqref{G-function} have known singularities as function of $E$: One sees
from Eq.~\eqref{recurrence} that $f_{n+1}(x)$ has a simple pole at $x=n$ and
from \eqref{G-function} we deduce that $G^q_\pm(x)$ has simple poles at $%
x=0,1,2,\ldots$\ The $x_n^\pm$, where $G_\pm^q(x)$ vanishes, must be located
in between these poles; the number varies between zero and two, but because
the poles are simple, there can be no adjacent intervals which are both
devoid of zeros or contain both two zeros. This feature follows for $%
\Delta\ll 1,g$ from Eq.~\eqref{recurrence}. The zeros of $G_\pm^q(x)$ are
therefore smoothly distributed between the equidistant poles, i.e. $%
x_{n+1}^\pm-x^\pm_n\simeq 1$ (for fixed parity). The distance between poles
of $\tilde{G}^q_\pm(E)=G^q_\pm(x(q,E))$ on the energy axis depends on $g$:
If $x_1^{\text{pole}}=n$, $x_2^{\text{pole}}=n+1$, we find for $E_2^{\text{%
pole}}-E_1^{\text{pole}}=2\beta=2\sqrt{1-g^2}$. If $g$ approaches the
critical value $g_c=1$ from below, the distance between the poles of $\tilde{%
G}^q_\pm(E)$ goes to zero and it follows $E^\pm_{n+1}-E^\pm_n\simeq 2\sqrt{%
1-g^2}$. This is the \textit{spectral collapse} observed in previous
numerical investigations of the two-photon QRM~\cite%
{Chen2012,twophoton,collapse,lo}, which behaves similar to the two-mode
model in this respect. At the critical point $g=1$, the spacing between
adjacent discrete energy levels vanishes and the spectrum becomes continuous
- accordingly no normalizable states exist anymore. The number of bosonic
excitations in low-lying states grows very fast in approaching the critical
coupling~\cite{collapse,lo}, which makes numerical diagonalization more and
more difficult. In our approach, the qualitative properties of the collapse
can be deduced very easily by a simple analysis of the singularity structure
of the $G$-functions.

\textsl{Exceptional spectrum}: Koc et al.~\cite{Koc} have obtained isolated
and doubly degenerate solutions for the QRM, the quasi-exact Juddian
solutions~\cite{Judd}.
We have explicitly excluded degenerate eigenstates in the derivation of the $%
G$-functions pertaining to the regular spectrum. It follows that the Juddian
solutions belong to the exceptional spectrum. It can be analyzed along
precisely the same lines as for the QRM~\cite{Braak}.
For special values of the model parameters $g$ and $\Delta $, there are
eigenvalues which do not correspond to zeros of $G_{\pm }^{q}(x)$; the
corresponding energy has the value $E^{\text{exc}}=2\beta (n+q)-1$.
The necessary and sufficient condition for the occurrence of this eigenvalue
is
\begin{equation}
f_{n}^{q}\left( x_{n}\right) =0,  \label{exc-cond}
\end{equation}%
which provides a condition on the model parameters $g$ and $\Delta $. They
occur when the pole of $G_{\pm }^{q}\left( x\right) $ at $x=n$ is lifted
because the corresponding numerator in Eq.~(\ref{G-function}) vanishes. The
conditions for all $n$ can be obtained from the recurrence Eq.~(\ref%
{recurrence}). For example, the conditions for $n=1$, respectively $n=2$
read,
\begin{equation}
\left( 2q+1\right) g^{2}+\frac{\Delta ^{2}}{4}-1=0,  \label{iso_1}
\end{equation}

\begin{eqnarray*}
&&\left( 4qg^{2}-6\beta ^{2}+4+\frac{\Delta ^{2}}{2}\right) \\
&&\left( 4q-4\beta ^{2}\left( 1+q\right) +\frac{\Delta ^{2}}{4}\right)
-8g^{2}q = 0
\end{eqnarray*}
The first condition was also obtained in Ref.~\cite{Zhangyz}. All
exceptional solutions corresponding to condition \eqref{exc-cond} are doubly
degenerate, because the pole is lifted both in $G_{+}^{q}(x)$ and $%
G_{-}^{q}(x)$, they correspond therefore to crossing points between states
with different parity in the spectrum.
As shown in Fig.~\ref{spectrum05}, the first crossing point at $g=0.696$
coincides with the value obtained from Eq.(\ref{iso_1}) for $\Delta =0.35$, $%
q=1/2$.

\section{Conclusions}

We have derived $G$-functions for the two-mode QRM by using ESS for each
Bargmann and parity index. The zeros of the $G$-functions $G_\pm^q(x)$
determine the regular spectra. Doubly degenerate eigenvalues are given by
the conditions for lifting a pole in $G_\pm^q(x)$ at integer values of $x$,
which can be given in closed form.
Moreover, the known pole structure of $G_\pm^q(x)$ allows us to derive the
shrinking average level spacing as one approaches the critical point $g=1$
(spectral collapse) analytically, whereas any numerical approach is bound to
fail in the region close to the critical point due to the presence of highly
excited states even at low energy.

Our result adds the two-mode QRM to a list of models having a $G$-function
with general structure,
\begin{equation}
G_{\pm }\left( x\right) =\sum_{n=0}^{\infty }f_{n}^{{}}(x)\left[ 1\pm \frac{%
\Delta }{F(g)\left( n-x\right) }\right] L_{n}(g),  \label{universal}
\end{equation}%
where $f_{n}^{{}}(x)$ is determined recursively. In the QRM~\cite{Braak},
\begin{eqnarray*}
F(g) &=&1, \\
L_{n}(g) &=&g^{n}, \\
E &=&x-g^{2}.
\end{eqnarray*}%
and $f_{n}^{{}}(x)$ is determined by Eq.~(5) in \cite{Braak}. For the
present two-mode QRM we have,
\begin{eqnarray*}
F(g) &=&2\sqrt{1-g^{2}}, \\
L_{n}(g) &=&\left( n+2q-1\right) !\left( \frac{v}{u}\right) ^{n}, \\
E &=&2\sqrt{1-g^{2}}(x+q)-1.
\end{eqnarray*}%
and $f_{n}(x)$ is determined by Eq.~(\ref{recurrence}) for each $q$. The
two-photon QRM~\cite{Chen2012} fits into this general scheme as well,
\begin{eqnarray*}
F(g) &=&2\sqrt{1-4g^{2}} \\
L_{n}(g) &=&\frac{\left[ 2\left( n+q-\frac{1}{4}\right) \right] !}{n!}\left(
\frac{v}{2u}\right) ^{n}, \\
E &=&2\sqrt{1-4g^{2}}\left( x+q\right) -\frac{1}{2},
\end{eqnarray*}%
where the Bargmann index $q=\frac{1}{4},\frac{3}{4}$, and $f_{n}(x)$ is
determined by Eq.~(41) in \cite{Chen2012}.
In terms of $q$, $f_n(x)$ reads,
\begin{widetext}
\begin{equation}
f_{n+1}^{(q)}=\frac{\left( 1+4g^{2}\right) \left( n+q\right) -\beta
^{2}\left( x+q\right) -\frac{\Delta ^{2}}{4\left( n-x\right) }}{4g\left(
n+q+\frac{3}{4}\right) \left( n+q+\frac{1}{4}\right) }f_{n}^{(q)}-\frac{%
f_{n-1}^{(q)}}{4\left( n+q+\frac{3}{4}\right) \left( n+q+\frac{1}{4}\right) }
\label{recure_2p}
\end{equation}
\end{widetext}

The list may be expanded by other related models in the future. All
qubit-cavity models possessing a compact $G$-function of the type (\ref%
{universal}) share two common properties: Isolated doubly degenerate
eigenstates with energy $x(E)=n$ and regular spectra given by zeros of a $G$%
-function which is a linear combination of two formal solutions. Both
properties have their root in the discrete $\mathbb{Z}_2$-symmetry present
in all models belonging to this class, rendering them integrable.

\textbf{ACKNOWLEDGEMENTS} This work was supported by National Natural
Science Foundation of China under Grant Nos. 11174254 and 11474256, National
Basic Research Program of China under Grant No. 2011CBA00103. D.B.
acknowledges partial support by the National Science Foundation of China,
Grant No. 11474250. He wishes to thank Stefan Kirchner and Huiqiu Yuan for
warm hospitality during his stay at the Center for Correlated Matter,
Zhejiang University, Hangzhou.

$^{*}$ Email:qhchen@zju.edu.cn

$^{\dagger}$ Email:daniel.braak@physik.uni-augsburg.de

\end{document}